\documentclass[a4,aps,prb,amsmath,amssymb,superscriptaddress, showpacs, twoside]{revtex4}

\input {epsf.sty}
\usepackage{graphicx}
\newcommand\ie {{\it i.e. }}
\newcommand\eg {{\it e.g. }}

\newcommand\cf {{\it cf. }}

\newcommand\noi{\noindent}
\newcommand{\ee}{\end{eqnarray}}
\newcommand{\pref}[1]{(\ref{#1})}

\newcommand\ket [1] {|#1 \rangle }

\newcommand\half{\frac 1 2 }

\newcommand{\pdd}[2]{{\partial{#1}\over\partial{#2}}}
\newcommand{\pa}[1]{\partial_{#1}}

\newcommand{\be}[1]{\begin{eqnarray} \mbox{$\label{#1}$} }

 \newcommand{\ok}{\bar k}
 \newcommand{\oz}{\bar z}
 \newcommand{\zd}{Z^{\dagger}}

 \newcommand\kp{\ell^{2}(\vec k \times \vec p)}
 \newcommand\kph{\frac {\ell^{2}} 2 (\vec k \times \vec p)}

\begin{document}
\title{Charge and Current in the Quantum Hall Matrix Model }
\author{T. H. Hansson, J. Kailasvuori, A. Karlhede}
\affiliation{Fysikum, Stockholm University,
AlbaNova University Center,
SE - 106 91 Stockholm, Sweden }

\date{\today}

\begin{abstract}
 We extend the  quantum Hall  matrix model to include couplings to
 external electric and magnetic fields.  The associated current suffers
 from matrix ordering ambiguities even at the classical level.  We
 calculate the linear response at low momenta -- this is unambigously
 defined.
 In particular, we obtain the correct fractional quantum Hall
 conductivity, and the expected density modulations
 in response to a weak and slowly varying magnetic field.
 These results show that the classical quantum Hall matrix models
describe important aspects of the dynamics of electrons in the lowest
 Landau level.
 In the quantum theory the ordering ambiguities are more severe;  
 we discuss possible strategies, but we have not been 
 able to construct a good density operator, satisfying
 the pertinent lowest Landau level commutator algebra.

\end{abstract}


\maketitle

 \section {Introduction}

 \noindent
 It is well established that the  low-energy effective Lagrangian for
 a quantum Hall (QH) system is a Chern-Simons (CS) gauge theory.
 For the simplest case of the Laughlin states with filling fraction
 $\nu = 1/(2m+1)$ it takes the form,
 \be{eff}
 {\cal L} = \frac k {4\pi} \epsilon^{\mu\nu\lambda}
 a_{\mu}\partial_{\nu}a_{\lambda} - \frac e {2\pi}
 \epsilon^{\mu\nu\lambda}
 A_{\mu}\partial_{\nu}a_{\lambda} \, ,
 \ee
 where $k=1/\nu$, and $A_{\mu}$ is the external
 electromagnetic potential. This Lagrangian (and its generalizations
 to other fractions,  multi-component systems {\em etc.})
 correctly describes the quantum Hall conductance, and  also
 the ground state degeneracies on higher genus surfaces and edge
 excitations\cite{zee}.
 One can also naturally incorporate quasiparticle currents and edge
 modes with appropriate quantum numbers.

 There are, however, important aspects of the QH liquids that
 are not incorporated in the CS description.
 For the Laughlin fractions described by \pref{eff} the quantization
 of the QH conductance corresponds to quantization
 of $k$, which  does not follow from any general
 principle but should be regarded as an input.\footnote{
   There are, however, microscopic
   derivations of \pref{eff}, relying on various mean field
   approximations, where the quantization of $k$  follows from
   a statistical transmutation\cite{zhang92}.
   }
 For hierarchical states, the parameter $k$ is
 replaced by a matrix ${\bf K}$ with quantized entries which again are
 input parameters. Furthermore, it is  difficult,
 within the effective CS framework, to account for the
 successes of the composite fermion model and to single
 out the experimentally prominent Jain fractions $\nu = n/(2pn\pm
 1)$\cite{jain1}.

 There have been several recent attempts to formulate effective
 theories for the composite
 fermions, based on various CS mean field theories\cite{recent} and
 in these approaches the constraint that keeps the electrons in  the
 lowest Landau
 level (LLL) is very important. On the level of the effective theories
 this is manifested
 in the commutation relation
 \be{ccom}
 [\rho_{\vec k}, \rho_{\vec p}]  = 2i \sin\left(\kph
 \right)e^{\frac {\ell^{2}} 2 \vec k
 \cdot \vec p } \rho_{\vec k+\vec p} \, , \ \ \ \ \ \ell =
 (\hbar/eB)^{1/2} \ ,
 \ee
 for the Fourier components of the LLL-projected density operator $\rho$
 \cite{girvin}
 and in a dipole form for the quasiparticle
 current operator.
 Neither of these properties is captured by  effective CS theories
 like \pref{eff}.
 For instance, from \pref{eff} we get  $\rho(x)=j^0 = \frac{1}{2\pi}
 \epsilon^{ij}\pa i
 a_{j} = \frac{1}{2\pi} b$, hence the commutator \pref{ccom} vanishes
 $[\rho_{\vec k}, \rho_{\vec p}]=-4\pi i \hbar \nu (\vec p\times \vec
 k)\delta(\vec k+\vec p) = 0 $.

 In view of the above, the recently proposed matrix model for the
 QH effect is intriguing\cite{suss1,poly1}.
 This model -- which can  be shown to be a
 noncommutative version of the effective CS theory \pref{eff}
 -- has several appealing features.
 The quantization of $k$ follows from basic principles\cite{suss1,nair1}
 and
 the specific values corresponding to the Laughlin fractions
 follow by requiring the underlying particles to be
 fermions, much as in the microscopic Chern-Simons Ginzburg-Landau
 approach\cite{zhang92}.
  The classical ground state of the matrix model is an
 incompressible homogeneous state at the
 Laughlin filling fractions, and the  model also
 reproduces several important aspects of
 fractional QH physics,
 such as the quasihole charge and statistics, and the
 presence of edge states.

 In spite of all this  the matrix models have so far not added anything
 to our understanding of QH
 physics -- the Laughlin states are very well
 understood -- primarily through the Laughlin wave functions, but also
 from  various Chern-Simons mean field theories.  The matrix models
 will  be of real physical interest only if they can be generalized to
 describe other states. As discussed above, an important challenge
 is to understand the Jain fractions, for which there
 are very good wave functions, and where the mean
 field theory explicitly involves ``effective fermions'' filling
 auxiliary higher Landau levels.

 We think that a generalized matrix model might  be useful, since in
 the existing matrix model it is enough to specify that the original
 particles are fermions in order to get the correct Laughlin
 fractions. In Laughlin's theory it is the combination of fermi
 statistics {\em and} LLL dynamics that determines the wave functions,
 so, by analogy, we hope that the matrix model encodes the correct
 LLL dynamics. There are already some indications that this is the
 case. For a vanishing non-commutativity parameter $\theta$
  it is obviously true by construction, and
 it has also been shown that a certain class of solutions to
 the finite-dimensional matrix model describes independent point
 particles moving in a stong magnetic field at distances larger than
 $\theta/\ell$ where $\ell$ is the magnetic length\cite{poly1}.
 It is, however, far from clear that the matrix model captures the
 full LLL dynamics in general, and  in particular
 for the incompressible states describing the Laughlin liquids.
 It is the purpose of this paper to study this question, and in
 particular to construct a conserved electromagnetic current
 that will allow us to calculate response functions and to see
 if the charge commutator algebra \pref{ccom} is satisfied.

 We proceed by constructing, in the classical matrix model,
 a particle density and a
 corresponding conserved current density which  are then  coupled
 minimally to the external electromagnetic
 potential. The densities are
 complicated matrix functions, and furthermore, they are not unique --
 the noncommutativity of the  matrices leads to an ordering ambiguity
 already at the classical level.

 Using the infinite dimensional matrix model, which  describes the
 Laughlin states, we   calculate  the response to weak
 perturbing fields using linear
 response. This amounts to calculating the ground state density, the QH
 response and the density response to
 weak and slowly varying  magnetic fields. We show that the correct QH
 results are obtained for any choice of matrix ordering in the definition
 of the densities.

 We then turn to the quantum theory and discuss the construction of the
 density and current operators.  There are several constraints on
 the correct construction: the density must satisfy the commutation
 relations \pref{ccom}, be properly normalized (\ie reproduce the correct
 ground state density for homogeneus states),
 and in the commutative limit it must reproduce the
 physics of non-interacting particles in the LLL. The problem of
 constructing a
  density and a corresponding
 conserved current density now involves both a matrix and
 a quantum ordering problem,  and is, in general, very involved.  
 

 Our aim is to make this paper fairly self-contained, and accesible to
 the condensed matter comunity, so we start with a short review
 of the QH matrix models in the next section. The construction of the
 classical conserved current and the corresponding Lagrangian is in
 section 3, and the calculation of the linear response in
 section 4. The quantization and the discussion of the charge
 commutation relations is in section 5. We conclude in section 6 with a
 summary of our results, a guess, and some open problems.
 Some of the material in this paper was presented in \cite{hakarl1}.


 \section{The Quantum Hall matrix model}
 The original version of the matrix model, proposed by Susskind,
 describing particles of charge $-e$ in a constant magnetic field
 $B_z=B$,\footnote{This choice, which we here adhere to,
 results in the one-particle
 quantum commutator  $[x,y]=+i\ell^2$ and states
 $\phi_n \sim {\bar z}^n \exp{-\frac{1}{4\ell ^2}|z|^2}$.
 More common  in the  condensed matter community  is to choose $B_z=-B$,
 in which case  $[x,y]=-i\ell^2$ and   $\phi_n \sim { z}^n
 \exp{-\frac{1}{4\ell ^2}|z|^2}$. }
  is
 given by the Lagrangian\cite{suss1}
 \be{mat}
 L_{0} = \frac {eB} 2 {\rm Tr} \{ \left( \dot{ X^{a}}  - i[X^{a},
 \hat a_{0}]_{m} \right) \epsilon_{ab}X^{b} +
 2\theta \hat a_{0}\} \, ,
 \ee
 where $X^{a}(t)$, $a=1,2$,  and $\hat a_{0}(t)$ are Hermitian matrices --
 the
 latter is a  Lagrange multiplier imposing a constraint.
 The area $\theta$ that enters $L_{0}$ is the
 non-commutativity parameter, see \pref{tvang} below, and
 in the classical model the average density is given by $2\pi \theta$.
 Quantization introduces $\hbar$ and the
 magnetic length $\ell = (\hbar /eB)^{1/2}$. Since
 $2\pi \ell^{2}$ is the area per state in the lowest Landau level,
 the filling fraction $\nu$ will be related to the dimensionless
 parameter $\theta/\ell^2$. For quantum Hall states, where $\nu$ is some
 odd denominator fraction, $\theta$ will be $O(\hbar)$ and
 sensitive to quantum corrections. In fact, as discussed
 briefly below, the relation  is
 $\nu^{-1} = \theta/\ell^2 + 1 $.

 The Lagrangian \pref{mat} is formally very similar to
 the Lagrangian for one particle moving in a strong perpendicular magnetic
 field
 $L_{1p}=\frac {eB} 2 \varepsilon _{ab} \dot x^{a} x^{b}$, where $x^{a}(t)$
 is the position of the particle. In fact, when $\theta = 0$,
 \pref{mat} reduces,  upon solving the constraint, to $N$ copies of
 $L_{1p}$ ($N$ being the dimension of the matrices $X^{a}$). We note
 that, when including a potential $V(\vec x)$, the equations of motion
 obtained from  $L_{1p}$ give the drift current $\dot x^{a}=\frac 1
 {eB} \epsilon ^{ab} \partial_{b} V(\vec x) $, but  not the cyclotron
 motion. This is as expected since $L_{1p}$ corresponds to infinite
 cyclotron frequency and hence vanishing cyclotron radius for finite
 velocity.

 As mentioned in the introduction, there is a close
 connection between the matrix model \pref{mat} and the CS theory
 \pref{eff} in that the matrix model is equivalent to a
 non-commutative generalization of the latter.  This is how Susskind
 arrived at \pref{mat}. He noted that in a  hydrodynamical
 description of the QH fluid one obtains an abelian CS theory with
 additional  {\it non-linear} terms. He then proposed the non-commutative
 abelian CS theory -- which to lowest nontrivial order in
 the non-commutativity parameter $\theta$ agrees with the
 non-linear abelian CS theory -- as the
 effective QH theory.

 Varying the Lagrangian \pref{mat} with respect to the multiplier
 $\hat a_0$ yields the constraint
 \be{tvang}
 [X^{1},X^{2}]_{m}=    i\theta \, ,
 \ee
 where the subscript "$m$" denotes a matrix, as opposed to a quantum,
 commutator.  Taking the trace of \pref{tvang} we see that
 $\rm {Tr} (X^{1} X^{2}) \neq \rm {Tr} (X^{2} X^{1})$ if $\theta \neq 0$.
  Thus the constraint can only be solved
 by infinite matrices and, as is well known from
 quantum mechanics,  the solution is essentially unique. The theory
 is
 hence rather trivial -- there is only one state, corresponding to an
 incompressible fluid with constant density $\rho = 1/2\pi\theta$. In order
 to
 find other solutions, one must modify the constraint \pref{tvang}.
 This can  be done either by introducing external sources\cite{suss1},
 or by introducing a new dynamical field that couples to
 $\hat a_{0}$\cite{poly1}. The latter approach has the advantage of using
 finite matrices, which means that all the ordinary rules for matrix
 manipulations, such as the cyclicity of the trace,  can be used.
 The finite matrix model, due to Polychronakos, is obtained by adding
 the following piece to the Lagrangian \pref{mat}
 \be{fin}
 L_{b} = \Psi^\dagger (i\partial_0 - \hat a_{0})\Psi \ \ ,
 \ee
 where $\Psi$ is a complex bosonic $N$-vector. The $X^a$'s in
 \pref{mat} are now Hermitian $N\times N$ matrices  and the constraint
 \pref{tvang} is  modified into
 \be{ptvang}
 [X^{1},X^{2}]_{m}=    i\theta - \frac i {eB}
 \Psi\Psi^\dagger \, .
 \ee
 Taking the trace gives $\Psi^\dagger \Psi = NeB\theta$, which
 allows for finite $N$ solutions also when $\theta \neq 0$.

 The Lagrangian
 $L=L_{0}+L_{b}$ is invariant under the $U(N)$ gauge transformations
 \be{ung}
 \Psi   &\rightarrow & U \Psi \nonumber \\
  X^a   &\rightarrow & UX^a U^{\dagger}  \nonumber \\
  \hat a_{0} &\rightarrow & U \hat a_{0} U^{\dagger}  -
  Ui\partial_{0}U^{\dagger}\ \ ,
 \ee
 where $U$ is a $U(N)$-matrix.

 This classical model was solved in  \cite{poly1}. Identifying
 the eigenvalues of
 the matrices $X^{1}$ and $X^{2}$ as the positions and momenta of the
 $N$ particles respectively, one finds that when the particles
 are far apart, compared to  $\theta/\ell$, they move as
 $N$ independent  particles in a strong perpendicular magnetic field
 (\ie governed by  $L_{1p}=\frac {eB} 2 \varepsilon _{ab} \dot x^{a}
 x^{b}$). On the other hand,   when a
 potential  $\propto (X^a)^2$ is added -- this attracts the particle
 to the origin -- the ground state is indeed
 a circular droplet with  constant bulk density $\rho   \sim
 1/2\pi\theta $. The excitation spectrum is that of the
 Calogero model --
 in particular this means that the low lying excitations can be
 interpreted as surface modes.

 The most interesting properties of the matrix models are appearant
 only after quantization. As usual for gauge theories, there are two
 ways to handle the constraint.
 In the first, which is similar to the  Gupta-Bleuler quantization of
 gauge theories, one  quantizes
 the full extended phase space to get the
 simple canonical commutation relations,
 \be{urcom}
 [X^{1}_{mn}, X^{2}_{rs}] &=& \frac {i\hbar}{ eB} \delta_{ms}\delta_{nr}
 \\
 {[\Psi_m , \Psi^\dagger_n ]}     &=& i\hbar \delta_{mn} \nonumber \, ,
 \ee
 and then implement the constraint \pref{tvang} or \pref{ptvang}
 as a projection operator to define
 physical states. We shall refer to this method
 as unconstrained  quantization.

 Alternatively, one proceeds like in Coloumb gauge quantization of a
 gauge theory, and first solves the ``Gauss' law'' constraint
 \pref{tvang} or  \pref{ptvang} in
 some  suitably choosen gauge (\eg  $X^{1}$
 diagonal). Writing the Lagrangian in terms of the physical variables
 one can then read off the canonical commutation relations.
 In this approach, there is no projection -- all states are
 physical --  but the commutators between the elements of the matrices
 become complicated. We  shall refer to this  method as
 constrained quantization.

 The finite matrix model also gives an unambigous
 relation between the filling fraction $\nu$ and the
 Chern-Simons level  $\theta/\ell^2$.
 This was originally derived by Polychronakos by
 mapping the matrix model onto the quantum Calogero model with
 a coupling constant $\nu^{-1}(\nu^{-1} - 1)$,
 where $\nu^{-1} = \theta/\ell^2 + 1$.
 From the known ground state of the Calogero model
 he could then identify $\nu$ with the filling
 fraction\cite{poly1}.
 The quantum shift in the
 relation between $\nu^{-1}$ and the level
 $\theta/\ell^2$ is due to the zero
 point motion of the particles, as explained
 by Hellerman and Susskind\cite{suss2}.
 Assuming the original particles to be fermions,
 one can furthermore show that the filling fraction
 is quantized to the Laughlin values $\nu^{-1} =
 2m + 1$.

 We should also mention that invariance under large gauge
 transformations requires the Chern-Simons level $k$
 to be an integer independent of
 the physical interpretation of the theory\cite{poly1,nair1}.
 For further discussion of
 the QH matrix models we refer to
 \cite{suss1,poly1,poly2,mor,heller}.


 \section{A generalized matrix model}
 We now generalize the matrix model \pref{mat} and \pref{fin} to include
 coupling to
 an external
 electromagnetic field, $A_\mu$. We use the more general finite
 dimensional model, $L=L_{0}+L_{b}$,  which has the additional advantage
 of allowing cyclic permutations in the trace. At the end of the
 section we comment on the infinite dimensional case -- making explicit
 the modifications this introduces. Since a large constant magnetic field
 $B$,
 perpendicular to the plane, is already incorporated in the model, we let
 $\delta B$ denote the magnetic field  corresponding to
 $A_\mu$ and assume that it has no constant component. To preserve gauge
 invariance, we must construct a conserved current from the matrix
 variables. It is useful to remember how to treat a single particle
 moving along the trajectory $\vec x(t)$. Here the particle
 and current  densities are given by $\rho (\vec y, t) = \delta(\vec y
 -\vec x(t))$ and $\vec j(\vec y, t) = \dot{\vec
 x}\delta(\vec y -\vec x(t))$,  respectively. Current conservation
 $\vec\nabla\cdot\vec
 j + \partial_t\rho =0$ then follows immediately using the chain rule.
 Using a $\delta$-function to define the particle and current densities
 corresponds to
 a point particle -- but in a non-relativistic theory, one can in fact
 use any profile function $f(\vec y -\vec x(t))$, corresponding to a
 rigid charge distribution moving with  velocity $\dot{\vec x}(t)$.

 In the matrix model, the diagonal elements of the
 matrices, $X^{a}_{mm}$, can, in a suitable gauge,  be interpreted
 as the guiding
 center coordinates for the particles when these are far apart
 compared to $\theta/\ell$\cite{poly1}.
 With this in mind it is natural to define the particle density
 in the matrix model as
 \be{charge}
 \rho(\vec y, t) &=&  {\rm Tr} [ \hat \delta (y^a  - X^a(t))] \ \ ,
 \ee
 where $\hat \delta (y^a  - X^a) $ is a matrix-valued kernel.
 With the same picture in mind, we make the following
 guess for the current density related to the guiding center motion,
 \be{current}
 \vec j(\vec y, t) &=&  {\rm Tr}[( \dot{\vec X}
 -i[\vec X, \hat a_{0}]_{m})\hat \delta (y^a -
 X^a(t))]  \ \ .
 \ee
 (In the presence of an inhomogeneous magnetic field,
 there is an additional, purely solenoidal, contribution to the current
 related to the uncancelled
 cyclotron motion of neighbouring electrons, which will be discussed
 below.)

 The densities \pref{charge} and \pref{current} are invariant under the
 $U(N)$-transformations \pref{ung} -- provided only that the kernel
 transforms covariantly: $\hat \delta
 \rightarrow U \hat \delta U^{\dagger}$.
 However,  since $X^{a}$ are non-commuting objects there are
 ordering ambiguities already at the classical level, leading to
 inequivalent classical theories. For a
 kernel of the form
 \be{gker}
 \hat \delta (y^a - X^a) = \int \frac{d^2k}{(2\pi)^2}\, f(k_a (y^a - X^a),
 \theta
 k^2) \ \ ,
 \ee
 where $f(z,x)$ is dimensionless and analytic
 in $z$, \pref{current} does indeed give a conserved current.
 The proof is easy:
 \be{prop}
 \partial_t {\rm Tr} [k_a (y^a - X^a) ]^n  &=& -{\rm Tr}
 \sum_{m=0}^{n-1}
 [k_b (y^b - X^b) ]^m  k_a \dot X^a [k_c (y^c - X^c) ]^{n-m-1}
 \nonumber \\ &= &
 -n{\rm Tr} k_a \dot X^a [k_b (y^b - X^b) ]^{n-1} = -
 \partial_{y^a}
 {\rm Tr} \dot X^a [k_b (y^b - X^b) ]^n  \nonumber
 \ee
 and
 \be{prop2}
 \partial _{y^a} {\rm Tr}\left( [X^a, \hat a_{0}] [k_b (y^b - X^b)
 ]^n\right) &=&
 n {\rm Tr}( [k_{a}X^a, \hat a_{0}] [k_b (y^b - X^b) ]^{n-1} \nonumber \\
 &=&
 -n {\rm Tr}( [k_{a}X^a, [k_b (y^b - X^b) ]^{n-1}] \hat a_{0}) = \,
 0 \ \ ,
 \ee
 where we used  the cyclic property of the trace.
 One kernel of the form \pref{gker} is
 \be{weylker}
 \hat \delta_W(y^a  - X^a) = \int \frac{d^2k}{(2\pi)^2}\, e^{ik_a
 (y^a - X^a)} g(\theta k^2) \ \ ,
 \ee
 where $g(0)=1$. The special case $g(x)=1$ is known as Weyl-ordering.
 The corresponding densities become, in the
 $\hat a_{0}=0$ gauge,
 \be{cc}
 \rho_{\vec k} &=& {\rm Tr}\left(e^{-ik_{a}X^{a}}\right)g(\theta k^2)
 \\
 j_{\vec k}^{a}  &=& {\rm Tr}\left(
 \dot{X^{a}}e^{-ik_{b}X^{b}}\right)g(\theta k^2) \, . \nonumber
 \ee
 Note the formal similarity to the one-particle densities.

 However, more general $O(2)$-invariant -- and $U(N)$-covariant --  kernels
 of the form
 \be{mgker}
 \hat \delta (y^a - X^a) = \int \frac{d^2k}{(2\pi)^2}\, f(k_{a} (y^a -
 X^a),
 \epsilon_{ab}k^{a} (y^b - X^b), \theta k^2) \ \
 \ee
 are possible. (The kernels are assumed to be Hermitian.)  For such
 kernels,
 \pref{current} does not generally give a conserved current. However, using
 \pref{charge} and current conservation it is
 straightforward to define a current as an expansion in the matrices $\dot
 X^{a}, \, k^{a} (y^a - X^a) $ and $\epsilon_{ab}k^{a} (y^b - X^b)$ that is
 conserved.
 An  important example  of this type is the anti-ordered kernel
 \be{aoker}
 \hat \delta_{\mathrm{ao}} (z-Z,\oz-Z^{\dagger}) = \int \frac{{\rm
 d}^2k}{(2\pi)^2}\,
 e^{\frac{i\ok}{2} (z-Z)} e^{\frac{i k}{2}(\oz- Z^{\dagger})}  \ \ ,
 \ee
 where  $z=y^{1}+i y^{2}$, $k=k^{1}+i k^{2}$ and $Z=X^{1}+i X^{2}$.
 It  turns out that the classical limit of the charge density operator in
 the  quantized
 matrix model discussed below is given by this kernel. The explicit
 expression for the corresponding conserved current in the $\hat
 a_{0}=0$ gauge is, using the cyclic property of the trace,
 \be{aocurr}
 j(z,{\bar z} )=j_{x}+ij_{y} =
 {\rm Tr}\left[ \dot Z \int \frac{{\rm d}^{2} k}{(2\pi)^2} \,
 e^{\frac{i}{2}(\ok \bar z +
 k\oz)}  \sum_{n=0}^{\infty}\sum_{m=0}^{n-1}
 \frac 1 {n!} (-\frac{i\ok}{2} Z)^{n-m-1}e^{-\frac{ik}{2} \zd}
 (-\frac{i\ok}{2} Z)^{m}\right] \, .
 \ee

 Having defined a particle density and a conserved current it is
 straightforward to couple the matrix model $L_{0}+L_{b}$ in \pref{mat}
 and \pref{fin}
 to an external electromagnetic field by adding
 \be{new0}
 L_{int} =  e\int d^{2}x \, [\rho (\vec x, t) A_{0}(\vec x, t) - \vec
 j (\vec x, t)\cdot \vec A(\vec x, t)] \ \ .
 \ee
 We can cast the resulting Lagrangian in a more transparent form by
 noting that
 the current always can be written in the form \pref{current}
 although in general the kernel defining the current will differ from
 the one defining the charge. For example, the  kernel defining
 the current $j$ corresponding to the anti-ordered charge kernel, can
 be read directly from \pref{aocurr} replacing the ordinary time
 derivative with the corresponding covariant derivative.
 The resulting Lagrangian is,
 \be{new}
 L &=& e{\rm Tr} \{ \left( \dot{ X^{a}}  - i[X^{a},
 \hat a_{0}]_{m} \right) \left(\frac {B}{2}\epsilon_{ab}X^{b} - \hat
 A_{a}\right) \\
 &+& B\theta \hat a_{0} +  \hat A_{0}
 -  \frac{\hbar}{2M} \delta \hat B       \}  +
 \Psi^{\dagger}(i\partial_{0}-\hat
 a_{0})\Psi \ \ . \nonumber
 \ee
 Here
 the matrices $\hat A_{\mu}(X^{a},t)$, $\mu = 0,1,2$, are related
 to the ordinary gauge potential using the kernel $\hat \delta_{(\mu)}$:
 \be{gendef}
 \hat A_{\mu}(X^{a},t) = \int  {\rm d}^{2}x\,
 A_{\mu} (x^{a},t)
   \hat\delta_{(\mu)}(x^a - X^a)  \, ,
 \ee
 and  the label $\mu$ on the
 kernel reminds us that it in general depends
 on which component of $A_{\mu}$ it multiplies.
 In \pref{new} we have also included a term $\sim \delta \hat
 B(X^{a},t)$, where $\delta \hat B$ is related to the magnetic field
 $\delta
 B(x^{a},t)$, caused by $A_{\mu}(x^{a},t)$, via $\hat \delta$ as in
 \pref{gendef}. $M$ is the effective  mass of the electrons.
 This is the Simon-Stern-Halperin ``magnetic moment'' interaction that
 encodes the variation in the Landau energy in a
 varying magnetic field\cite{halp1}.
 It gives an additional solenoidal term in
 the current -- due to the non-cancellation of the cyclotron
 motion currents -- that is not included in \pref{current}.

 Note that
 the charge and current density operators derived from \pref{new} do
 not have an explicit dependence on $\Psi$, but that there is an implicit
 dependence since  $\Psi$ enters the constraint and thus determines
 the form of the solutions for the matrices $X^{a}$.

 It is clear by inspection that  the Lagrangian \pref{new} is
 invariant  under the noncommutative gauge transformation \pref{ung},
 and the symmetry under the usual electromagnetic gauge transformation
 $\delta A_{\mu}(x) =
 \partial_{\mu}\Lambda(x)$ is ensured by construction since the
 corresponding current is conserved. For general kernels the gauge
 variation of the matrices $\hat A_{\mu}$ is rather complicated, but
 for the special case of Weyl ordering it takes a simple form very
 reminiscent of the commutative case.
 \footnote{
      Using \pref{weylker}, $\delta A_{\mu}(x) = \partial_{\mu}\Lambda(x)$
      induces the following gauge variations in the matrices
     $
      \delta \hat A^{a}_{mn} =
      - \left({\partial \hat \Lambda }/
      {\partial X^{a}}\right)_{mn} =
       -  {\partial {\rm Tr} \hat \Lambda} / {\partial X^{a}_{nm}}$
       and $ \delta \hat A_{0} = \pa 0 \hat \Lambda \, $
      where   the matrix-valued function $\hat\Lambda (X^{a},t)$ is
     related to $\Lambda(x^{a},t)$ using the  Weyl kernel \pref{weylker} in
     \pref{gendef}, and the derivative  satisfies  $\partial e^{-ik_{b}X^{b}}/\partial X^{a} =
       -ik_{a}e^{-ik_{b}X^{b} }$.                             }

 In the extended model \pref{new}, the constraint corresponding to
 \pref{ptvang} is modified and  depends on $A_{a}$:
 \be{ntvang}
 [X^{a},\epsilon_{ab}X^{b} - \frac 2 {B} \hat A_{a} ]_{m}=
 2i\theta - \frac {2i} {eB} \Psi \Psi^{\dagger} \, ,
 \ee
 and we note that contrary to \pref{ptvang} this does not determine the
 commutator between $X^1$ and $X^2$. However, if we pick the gauge
 $\hat A_{2} = 0$, then  we can rewrite the constraint as
 \be{cntvang}
 [X^{1}, X^{2} - \frac 1 {B} \hat A_{1} ]_{m}=
 [X^{1}, \frac 1 {eB} P^{1} ]_{m}= i\theta -\frac {i} {eB} \Psi
 \Psi^{\dagger}\, ,
 \ee
 where $P^{1}$ is the momentum conjugate to $X^{1}$ as calculated from
 the Lagrangian \pref{new}. Clearly, the above construction can be
 carried out in any linear gauge, but we do not know how to
 handle a general gauge.
 Note that the simple commutation relation \pref{cntvang} is between
 $X^{1}$ and
 $P^{1}=eBX^{2}-e\hat A_{1}$,  not between $X^{1}$ and $X^{2}$, and
 this makes it hard to evaluate  the densities \pref{charge} and
 \pref{current}. However, as
 long as we are interested
 only in linear response we can expand the expression for the density as
 will be shown in the next section.
 That the constraint becomes simple when expressed in canonically
 conjugate variables, is in fact very important since it assures that
 the arguments given in \cite{suss1} and \cite{poly1} for the relation
 between the parameter $\theta$ and the statistics of the underlying
 particles hold true also in the extended theory \pref{new}.

 We now comment on the case of the infinite dimensional matrix
 model \pref{mat} and its coupling to an electromagnetic field.
 The above analysis used the  cyclicity of the trace --
 \eg in \pref{prop2} and \pref{aocurr} -- which does not hold for
 infinite matrices.
 Also, defining the particle density  as in
 the finite dimensional case, \ie by  \pref{charge}
 with the kernel still given by  \pref{mgker},
 we can due to the lack of cyclicity
 no longer use expressions like \pref{current} or \pref{aocurr} for the
 conserved currents. It is however straightforward to construct the
 conserved
 currents by carefully keeping track of the matrix ordering.
 So will for instance the current corresponding to the Weyl-ordered
 kernel \pref{weylker} take the form,
 \be{infw}
 j^a(\vec x) = \int \frac{{\rm d}^{2} k}{(2\pi)^2} e^{ik_bx^b}
 {\rm Tr}\left[  \sum_{n=0}^{\infty}\sum_{m=0}^{n-1}
 \frac 1 {n!} (-ik_cX^c)^{n-m-1}{\dot X}^a (-ik_dX^d)^{m}\right] \, .
 \ee
 Other kernels give similar expressions.
 It is now clear that we cannot rewrite the interacton
 Lagrangian \pref{new0} of the
 generalized infinite matrix model on the simple form \pref{new}.
 Instead the coupling to the external vector potentials will take
 the more complicated form,
 \be{Lcoup}
 L_{\vec A} = -e\int d^2x\, A_a(\vec x ,t)
 \int \frac{{\rm d}^{2} k}{(2\pi)^2} e^{ik_bx^b}
 {\rm Tr}\left[  \sum_{n=0}^{\infty}\sum_{m=0}^{n-1}
 \frac 1 {n!} (-ik_cX^c)^{n-m-1}(D_0 X^a) (-ik_dX^d)^{m}\right] \, ,
 \ee
 \cf \pref{aocurr} and \pref{new0},
 where $D_0 X = \dot X - i[X, \hat a_0]$ is the covariant time
 derivative.
 Varying with respect to $\hat a_{0}$ one however
 finds that the constraint
 still has the form \pref{ntvang} (with $\Psi=0$), where $\hat
 A_{\mu}$ is given by \pref{gendef} and the kernel $\hat
 \delta_{(\mu)}$ is the same as for finite matrices. The point is
 that although trace manipulations using cyclicity are not allowed
 on the level of the Lagrangian, they are permissible in the
 equations of motion if we (as customary) assume
 that the {\em variation} $\delta \hat a_{0}$ of the
 multiplier matrix has compact support\cite{harvey2001}.

 In the simplest case when the constraint takes the form
 $[X^1,X^2]_m=i\theta$,
 then \pref{aoker} -- and probably any sensible kernel
 \pref{mgker} --
 can be written as a generalized Weyl kernel \pref{weylker}. Also,
 if one assumes the Hamiltonian to be a scalar potential, then
 by using the equations of motion to replace
  $\dot X_i$ in \pref{infw} by an expression in $X_j$'s,
  one can see that all the considered kernels
 $\hat \delta_{(\mu)}$  are in the infinite case
  related by factors of the type $g(\theta k^2)$ with $g(0)=1$.
 This observation will be useful in the next section.


 \section{Linear Response in the classical  matrix model}

 In this section we calculate the response of the infinite dimensional
 classical
 matrix model to electromagnetic perturbations in the linear approximation.
 We
 remind the reader that this model describes only one state with
 constant density $\rho=1/2\pi\theta$ -- nevertheless we obtain non-trivial
 results.

 We shall consider three quantities, the quantum Hall response to
 a constant electric field, the ground state density,
 and the response to a weak, static but slowly modulated
 magnetic field. The two first examples involve response at wave vector
 $\vec k = 0$
 and the last at small $\vec k$, \ie $k^2\theta \ll 1$. From the above
 discussion of the
 infinite matrix model it follows that the classical ordering ambiguities
 will be of no
 relevance for these quantities since they amount to correction terms $\sim
 k^2\theta$.
 For simplicity, we use the generalized
 Weyl-kernel \pref{weylker}, but any choice of the form \pref{gker} would
 do.

 \subsection{The quantum Hall response}
 We calculate the current response to an applied constant electric
 field of the form
 $A_{0} = -E_{a}x^{a}$.
 To evaluate the current, we first need to determine $\dot {X^{a} }$
 from
 the equations of motion, a problem  very similar to the QH
 droplet in a quadratic matrix potential
 considered by Polychronakos\cite{poly1}. Varying the Lagrangian with
 respect to $X^{a}$, yields\footnote{Here we  again  the
 cyclicity of the trace  in the  equations of motion.
 This assumption can be avoided for a kernel that behaves like a
 $\delta-$function:  $\int {\rm d}^2x\,  x^a \hat \delta (x,X) =
 X^a$ -- then we have $\hat A_{0}=- E_aX^a$. Substituting this into the
 Lagrangian and varying with respect to $X^{a}$ gives \pref{eom}.}
 \be{eom}
 B \dot {X^{a}} = \epsilon^{ab}  E_{b}{\bf 1} \, .
 \ee
 Because of $\dot {X}^{a} \propto{\bf 1}$,
 it follows from expressions
 like \pref{cc} or from the discussion at the end of section III,
 that the electromagnetic current density  is given by
 \be{qhe}
 J^{a}_{\vec k} =-e j^{a}_{\vec k} =
 -\epsilon^{ab} \frac {eE_{b}} B \rho_{\vec k} \, ,
 \ee
 and for  $\rho=\nu /2\pi\ell^2=\nu eB/h$
 a Fourier transformation yields
 \be{qhe2}
 J^{a} = -  \nu \frac {e^{2}} h \epsilon_{ab} E^{b} =
 -\sigma_{H} \epsilon^{ab} E_{b} \, .
 \ee
 >From refs. \cite{suss1,poly1} we know that if the particles
 described by the matrix model are fermions, then $\nu^{-1}
 =2m +1$ and \pref{qhe2} gives the Laughlin values
 for the quantized Hall conductance $\sigma_{H}$.

 \subsection{The ground state density}
 The response to a constant external electric
 potential $\delta A_{0}$ will  simply give the density of the system
 \be{gse}
 \rho(\vec x) = \pdd L {A_0} = {\rm Tr} \, \hat \delta(x^a-X^a) \, .
 \ee
 In the finite case this is difficult to calculate because of the
 presence of $\Psi$ in the constraint, but in the infinite case
 the calculation can be
 conveniently done by using coherent states.   Defining
 eigenstates of the lowering operator
 \be{low}
 a\ket \alpha = \alpha\ket\alpha \ \  , \ \ \ \ {\rm where} \ \ \ \
 a = \frac 1 {\sqrt{2\theta}} (X^1 + iX^2) \ \ \ \ ,
 \ee
 and using the commutator $[a,a^{\dagger}]=1$  appropriate for the infinite
 matrix model in the absence of external fields, one has
 \be{cotr}
 {\rm Tr} : {\cal O} (a,a^\dagger): \ \ =
 \int \frac{{\rm d}\alpha {\rm d}\bar{\alpha} }{2\pi i}
 {\cal O} (\alpha,\bar{\alpha}) \ \ ,
 \ee
 where : : indicates normal ordering of the operator
 ${\cal O} (a,a^\dagger)$.

 For  the Weyl-ordered  kernel \pref{weylker} ($g=1$) we now have,
 \be{calc}
 {\rm Tr} e^{-ik_aX^a} &=& {\rm
 Tr}e^{-i\sqrt{\frac{\theta}{2}}(\overline k a + k a^\dagger)} =
 e^{\frac{ \theta |k|^2}{4}} {\rm Tr}
 e^{-i\sqrt{\frac{\theta}{2}}kZ^\dagger}
 e^{-i\sqrt{\frac{\theta}{2}}\overline k Z}= \\
 &=&e^{\frac{ \theta |k|^2}{4}}
 \int\frac{{\rm d}\alpha {\rm d}\bar{\alpha} }{2\pi i}
 e^{-i\sqrt{\frac{\theta}{2}}(k\overline\alpha + \overline k \alpha)}
 =
 e^{\frac{ \theta k^2}{4}}
 \int\frac{{\rm d}^2\alpha }{\pi }
 e^{-i\sqrt{2\theta}k_i\alpha^i}=
 e^{\frac{ \theta k^2}{4}}
 \frac {2\pi}{\theta}\delta^2(-\vec k) \nonumber \ \ ,
 \ee
 where $k=k_1 + ik_2$. Substituting \pref{calc} in \pref{weylker}
 and \pref{gse} we obtain
 the constant density
 $\rho = \frac 1 {2\pi\theta}$, as expected.

 Notice that the  factor $e^{\frac{ \theta k^2}{4}}$
 does not  affect the
 result because of $\delta^2(\vec k)$. For the same reasons it follows
 from the discussion at the end of
 section III  that the result is independent of the choice
 of kernel.

 \subsection{Response to a weak perturbing B field}
 It is in general hard to calculate the response to an external
 magnetic field since, as explained above, \pref{cntvang}
 only gives an implicit relation for
 the matrix $\hat A^{a}(X)$. It can however be done for a
 weak, and slowly varying perturbing field of the form
 $\delta B(\vec x) = \epsilon B \sin (\vec q\cdot\vec x)$.

 In the $A^2 =0$  gauge we have $A^1 =\frac{\epsilon B}{q_2}
 \cos (\vec{q}\cdot\vec{x}) $, and  the
 constraint \pref{cntvang} (with $\Psi =0$). The annihilation
 operator then becomes
 \be{}
 a = \frac 1 {\sqrt{2\theta}} (X^1 + i(X^2-\frac{1}{B} \hat{A}^1))\, ,
 \ee
 and $k_aX^a =\sqrt{\frac {\theta} 2}(\bar{k}a+ka^\dagger)
 + \frac{k_2}{B}\hat{A}^1$. By expanding in $\epsilon$ and $\theta k^2$, we
 obtain
 \be{}
 \rho_k = {\rm Tr} e^{-ik_aX^a}= {\rm Tr}[
 e^{-i\sqrt{\frac{\theta}{2} }(\bar{k}a+ka^\dagger) }
 (1-\frac{ik_2}{B}\hat{A^1} +O(\epsilon^2) )
 (1+\epsilon O(\theta k^2))]  \, .
 \ee
 To  evaluate
 \be{drho}
 \delta \rho_k = -{\rm Tr}[
 e^{-i\sqrt{\frac{\theta}{2} }(\bar{k}a+ka^\dagger) }
 (\frac{ik_2}{B}\hat{A^1} +\epsilon O(\theta k^2)+ O(\epsilon^2) )]\, ,
 \ee
 we first write
 \be{something}
 \hat{A}_1(X)= \frac{\epsilon B}{2q_2}[e^{iq_aX^a}+e^{-iq_aX^a}]
 =\frac{\epsilon B}{2q_2}
 [e^{i\sqrt{\frac{\theta}{2} }(\bar{q}a+qa^\dagger)}+
 e^{-i\sqrt{\frac{\theta}{2} }(\bar{q}a+qa^\dagger)}+
 O(\epsilon)] \, ,
 \ee
 and then calculate the trace
 with the same method as in subsection B above to get
 \be{}
 \delta \rho_k &=&
 -\frac{i\epsilon k_2}{2q_2}
 {\rm Tr}[e^{-i\sqrt{\frac{\theta}{2} }(\bar{k}a+ka^\dagger) }
  [e^{i\sqrt{\frac{\theta}{2} }(\bar{q}a+qa^\dagger)}+
 e^{-i\sqrt{\frac{\theta}{2} }(\bar{q}a+qa^\dagger)}]]
 +\theta^{-1}O(\epsilon^2 )+\theta^{-1}O(\epsilon \theta k^2 ) \\
 &= &-
 \frac{i\pi\epsilon}{\theta}
 \left[
 \frac{k_2}{q_2}\left(\delta^2 (\vec{q}-\vec{k})+
 \delta^2 (\vec{q}+\vec{k})\right) +
 O(\epsilon )+O( \theta k^2 )\right]\, .
 \nonumber
 \ee
 Taking the Fourier transform, we obtain finally
 \be{res}
 \delta \rho(\vec x) =
 \frac \epsilon {2\pi\theta} \left[ \sin (\vec q\cdot\vec x)
 +O(\epsilon)+O(\theta k^2)\right ] \, ,
 \ee
 which is  the density response expected in a quantum Hall
 system.

 \section{Charge and current in the quantum matrix model}
 As discussed in the introduction, the Fourier components of the
 density operator projected onto the LLL satisfy the non-trivial
 commutation relation \pref{ccom}. Since the hope is
 that the QH matrix model captures the physics of the LLL,
 we would expect the correctly defined
 density operator  to satisfy \pref{ccom}.
 First notice that in the infinite matrix model, we can easily
 calculate
 the poisson brackets of the charge operators \pref{cc}, and turning
 these
 into commutators we get to lowest order in $\theta$ and $\hbar$,
 \be{lowest}
 [\rho_{\vec k}, \rho_{\vec p}]  = i \kp \rho_{\vec k+\vec p} +
 O(\hbar\theta , \hbar^{2}) \, .
 \ee
 This is at first hand quite surprising, since we would expect the
 $\theta = 0$ result to be that of the commutative Chern-Simons theory
 quoted in the introduction, $[\rho_{\vec k}, \rho_{\vec p}]=0$.
 If we however remember that the infinite matrix model only describes
 a single state, with density $\rho_0 =  1 /{2\pi\theta}$, we
 immediately  reproduce this result by substituting  \pref{calc}
 into the RHS of \pref{lowest}.
 Since the infinite matrix model does not allow any
 density fluctuations, we concentrate on the finite model from now on.

 At this point we face a serious operator
 ordering problem.
 Note that although the charge as defined in \pref{charge} with a
 given kernel \pref{mgker}
 defines a definite ordering of the products of operator-valued matrix
 elements when the kernel is expanded
 in a power series, any quantum reordering
 of the operators would correspond to the same classical matrix form,
 so the question of the correct quantum ordering of \pref{mgker} arises.
 We therefore have ordering ambiguites on two levels,  on
 matrix  as well as on quantum  level.

 If we quantize after solving the constraint, we know the correct
 answer in the $\theta = 0$ limit, since
 we  can then use commuting diagonal matrices
 where the entries  are simply the coordinates
 $z_n=x_n+iy_n$ and $\bar z_{n}$ of the $N$ independent particles,
 which satisfy  the LLL commutator $[z_m,\bar
 z_n]=2\ell^2 \delta _{mn}$.
 The anti-ordered operator
 \be{expao}
 \rho_k^{\mathrm{ao}}  =
 {\mathrm{Tr}}
 \, e^{-\frac{i\bar  k}{2} Z}   e^{-\frac{i k}{2} Z^\dagger  }
 \quad  {}_{\longrightarrow}^{\theta=0} \quad
 \sum_{n=0}^N e^{  -\frac{i\bar  k}{2} z_n }
  e^{-\frac{ik}{2} {\bar z}_{n} } \ \ ,
 \ee
 where $Z=X^{1}+i X^{2}$,  obeys \pref{ccom} for $\theta=0$.
 This shows that for $\theta =0$ the anti-ordered
  matrix kernel \pref{aoker}
 gives the correct quantum ordering. Unfortunately, this is not
  true for $\theta \neq 0$  --   series expansion shows that
 $\rho_k^{\mathrm{ao}} $ no longer satisfies \pref{ccom}.
 

Below we review an attempt to construct an operator that satisfies the LLL 
comutator algebra. The basic observation is that  for $N=2$, the known result \pref{expao} 
can be expressed in terms of quadratic operators which can be generalized to the 
$\theta \neq 0$ case but  satisfy a $\theta$-independent
algebra.  In the first version of this paper we erroneously claimed to have an expression in these
quadratic operators  that satisfied the LLL algebra \pref{ccom}. This conclusion was due to a 
logical error as will be pointed out.

 \subsection{Constrained quantization}
 Here we follow Polychronakos\cite{poly1} and use the matrices
 \be{conmat}
  X^1_{mn} &=& x_m \delta_{mn} \nonumber \\
 X^2_{mn} &=& y_m \delta_{mn} -\frac {i\theta} {x_m-x_n}
 (1-\delta_{mn}) \, ,
 \ee
 which  solve  the  constraint \pref{ptvang} in the gauge $\Psi =
 \sqrt {eB\theta} (1,1,\ldots1 )$. Quantizing, the
 diagonal matrix elements become canonically conjugate operators,
 satisfying (in complex notation),
 $[z_m,\bar  z_n]=2 \ell ^{2} \delta_{mn}$. Note that \pref{conmat}
 shows that when the particles are far apart $|\Delta z| \gg \theta /
 \ell$, the matrix model describes $N$ independent particles in the
 lowest Landau level.
 It is also clear that the ordering problem is very hard since
 $Z$ is a complicated, $\theta$-dependent, operator-valued matrix.

 It is not  difficult to find {\em some} operator that
 satisfies
 \pref{ccom} -- in fact $\rho_k =e^{  -\frac{i\bar  k}{2} {\rm Tr} Z }
  e^{-\frac{ik}{2} {\rm Tr} Z^\dagger }$  will do. However,
 this is nothing but the center of mass density
 operator, and will not define a good local particle
 density except in the trivial case
 of $N=1$.
 We shall demand that in addition to satisfying the quantum commutator
 algebra, the density operator must also reduce to
  the right-hand side of \pref{expao}
  in the limit
 $\theta =0$.
 As advertised above, we have found a solution to this problem for
 $2\times 2$ matrices  which is already  quite
 non-trivial, and we now present that construction.

 First we decompose the complex $2\times 2$  matrix $Z=X_1+iX_2$ as
 \be{decom}
 Z &=& \zeta{\mathbf{1}} + z \sigma^3 + i\vartheta \sigma^2 \nonumber \\
 Z^\dagger &=& \bar \zeta{\mathbf{1}} + \bar  z \sigma^3-
 i\vartheta\sigma^2 \ \ ,
 \ee
 where $\sigma^i$ are the Pauli matrices, $\vartheta = \theta/(z+\bar
 z)$, and where we have introduced  center of mass and relative
 coordinates,
 \be{zeta}
 \zeta &=&\half ( z_{1}+ z_{2})=
 \half{\mathrm{Tr}} \, Z \nonumber  \\
 z &=&\half (z_{1} - z_{2}) \ \ ,
 \ee
 satisfying $[\zeta,\bar \zeta]=[z,\bar  z]=\ell^2$,
 while all other commutators vanish.

 For $\theta = 0$ we expand the matrix exponentials in \pref{expao}
 using the explicit expresions \pref{decom} and the properties
 of the Pauli matrices to get
 \be{exp}
 \rho_k^{\rm ao}|_{\theta=0} =
 {\mathrm{Tr}}\, e^{-\frac{i}{2}\bar  k Z}
 e^{-\frac{i}{2} k Z^\dagger}   |_{\theta=0}=
 2\sum_{m,n=0}^\infty
 \frac {    ( \frac{-i\ok}{2} ) ^{2m}   }{   (2m)!   }
  \frac {   (  \frac{-ik}{2}  ) ^{2n}   } {  (2n)!   }
 e^{ -\frac{i\ok}{2}\zeta   }
 \left[      (z^2)^m (\oz^2)^n -\frac{|k|^2}{4}
 \frac{     (z^2)^m z\oz (\oz^2)^n  }{  (2m+1)(2n+1)   } \right]
 e^{-\frac{ik}{2}\bar\zeta}\, .
 \ee
 Note that since the particles are identical only even powers of the
 relative coordinate can appear in \pref{exp}.

 Next define the  quadratic operators,
 \be{ab}
 A  \equiv &\frac{1}{2}{\rm Tr} Z^2  -
 (\frac{1}{2}{\rm Tr} Z )^2  =&  z^2 - \vartheta^2 \nonumber \\
 \bar  A  \equiv &\frac{1}{2}{\rm Tr} (Z^\dagger )^2 -
 \frac{1}{4}({\rm Tr} Z^\dagger )^2  =&  {\bar  z}^2 - \vartheta^2 \\
 B \equiv&\frac{1}{2}{\rm Tr} ZZ^\dagger  -
 \frac{1}{4}{\rm Tr} Z {\rm Tr} Z^\dagger  =&
   z\bar  z + \vartheta^2 \,  . \nonumber
 \ee
 It is easy to verify  that they commute  with $\zeta$
 and satisfy  the following $\theta$-independent algebra,
 \be{qcom}
 [\zeta, \bar  \zeta ] &=&    \ell^2  \nonumber \\
 {[A,B]     }                &=&  2  \ell^2 A  \nonumber \\
 {[\bar  A,B] }         &=&  -2 \ell^2 \bar  A  \nonumber \\
 {[A,\bar  A]}            &=&  4 \ell^2 B -2\ell^4 \ \ .
 \ee
 This allows us to  define a $\theta$-dependent
 density operator by  substituting $z^2 \rightarrow A$,
 ${\bar  z}^2 \rightarrow \bar  A$ and
 $z\bar  z \rightarrow B $ in the  series expansion  \pref{exp} to get,
 \be{expanders}
 \rho_k =
 2\sum_{m,n=0}^\infty
 \frac {    ( \frac{-i\ok}{2} ) ^{2m}   }{   (2m)!   }
  \frac {   (  \frac{-ik}{2}  ) ^{2n}   } {  (2n)!   }
 e^{ -\frac{i\ok}{2}\zeta   }
 \left[      A^m {\bar A}^n -\frac{|k|^2}{4}
 \frac{     A^m B {\bar A}^n  }{  (2m+1)(2n+1)   }
 \right]
 e^{-\frac{ik}{2}\bar\zeta} \, .
 \ee
 By construction the commutator algebra of  the $\rho_{k}$ is
 $\theta$-independent, so the commutator \pref{ccom} can in 
 a $\theta$-independent way be reduced to a sum of terms of the
 form $A^mB^k{\bar A}^n$. We can however not reduce these terms 
 to sums of terms of the type $A^mB{\bar A}^n$ or $A^m{\bar A}^n$ 
 without using a $\theta$-dependent relation, and thus not 
 conclude that 
  since  \pref{exp} satisfies
 the algebra \pref{ccom}, so does  \pref{expanders}.
 This was the logical error referred to above.

 We now prove that the classical limit of $\rho_{k}$ in \pref{expanders}
 is the anti-ordered classical operator in \pref{expao} also for
 finite $\theta$.
 A general $2 \times 2$ matrix $Z$ with \emph{commuting} elements
 satisfies:
 \be{cetwo}
 0=Z^2 -Z{\rm Tr} Z+
 \frac{1}{2}\left({\rm Tr}^2 Z-{\rm Tr} Z^2\right){\mathbf 1}=
 (Z-\zeta{\mathbf 1} )^{2} - A{\mathbf 1}
 \ee
 where $\zeta$ and $A$ are defined in \pref{zeta} and \pref{ab}.
 Using this relation, we can rewrite the \emph{classical} anti-ordered
 operator,
 \be{rewao}
 \rho_k^{\mathrm{ao}}=
 {\mathrm{Tr}} \, e^{-\frac{i \bar  k}{2} Z} e^{-\frac{i k}{2}
 Z^\dagger} =
 {\mathrm{Tr}}
 e^{-\frac{i \bar  k}{2} \zeta} e^{-\frac{i \bar  k}{2}(Z - \zeta{\mathbf
 1}) }
 e^{-\frac{i  k}{2}(Z^{\dagger} - \bar\zeta{\mathbf 1}) }
 e^{-\frac{i k}{2} \bar\zeta } \ \ .
 \ee
 After expanding the matrix exponentials and using \pref{cetwo} to rewrite
 all even powers of $(Z - \zeta{\mathbf 1})^m
 (Z^{\dagger} - \bar\zeta{\mathbf 1})^n$ in terms of $A$, $\bar A$ and $B$,
 it is a matter of working out a few traces to recover
 the expansion \pref{expanders}.

 \subsection{Unconstrained quantization}
 In this approach the matrix elements of $X^a$ satisfy the quantum
 commutation
 relations \pref{urcom}, and the parameter $\theta$ enters only via
 the $U(1)$ part of the
 constraint \pref{ptvang} that defines the physical states.
 Since the density operator is an observable it must be $U(N)$
 invariant, and since
 the commutation relations \pref{urcom} preserve this invariance, it
 follows that
 the commutator of two densities must again be $U(N)$ invariant.
 (Note that since \pref{expanders}
 is expressed only in traces it is
 a manifestly $U(N)$ invariant expression, althouth
 it was derived in a particular gauge.)
 This would seem to imply
 that there could be no $\theta$-dependence in the operators, but
 this conclusion does not necessarily follow since
 we are  considering operators on an extended
 phase space. The
 commutator \pref{ccom} might have additional terms proportional to
 the constraint
 \pref{ptvang}, and this would still give the correct algebra on the
 physical subspace.
 However, since the gauge constraint explicitly involves this boundary
 field $\Psi$,
 the density operators can have a non-trivial $\theta$-dependence only
 if they also depend exlicitly on $\Psi$, and this would
 bring  us out of the class of operators defined by
 quantum reorderings of \pref{charge}. Although we shall not consider
 this possibility in this paper, it cannot be excluded {\it a priori}.

 Thus restricting ourselves to $U(N)$ invariant combinations of the
 matrices $Z$ and $Z^\dagger$, satisfying $[Z_{kl},Z_{mn}]=0$,
 $[Z_{kl},Z_{mn}^\dagger]=2\ell^2\delta_{kn}\delta_{lm}$, we note that
 the classical derivation of \pref{expanders} based on \pref{rewao}
 holds true also in the quantum theory since it did not involve any
 reordering of $Z_{kl}$'s and $Z_{mn}^\dagger$'s.
 (In the constrained
 quantization scheme this derivation brakes down since elements
 of the matrix $Z$ do not commute with each other.)
 The operators $A$, $\bar A$, $B$ and $\zeta$ are here  again
 defined in terms of the matrices by \pref{zeta} and \pref{ab},
  but
 they do not satisfy the algebra \pref{qcom}. Remember that the
 constrained and unconstrained $Z$ are not unitarily equivalent -- in
 that case they would have to satisfy the same algebra -- but
 related via a projection onto the subspace defined by the constraint.
 The new algebra is however very similar to \pref{qcom}, the only
 difference being the commutator
 \be{cvar}
 {[A,\bar  A]}            &=&  4 \ell^2 B -6\ell^4  \, .
 \ee
 This difference can be absorbed by a quantum
 reordering of \pref{expanders} in which $B$ is replaced by
 \be{btilde}
 \tilde{B} = \frac{1}{4}{\rm Tr} (ZZ^\dagger +Z^\dagger Z)  -
 \frac{1}{4}{\rm Tr} Z^\dagger {\rm Tr} Z = B-\ell^2 \ \ .
 \ee
 This gives ${[A,\bar  A]} = 4\ell^2 \tilde{B} -2\ell^4$,
 while leaving the other commutators unchanged
 (except for $B \rightarrow \tilde B$).
 Since the operator $B$ has the form of a trace of the
 number operator, it is not surprising that it differs in the
 two schemes since the number of degrees of freedom are $N$ and
 $N^{2}$ respectively. Note, however, that if the gauge-fixed $Z$'s are
 used, then $\tilde{B} =B$, so \pref{expanders} with $\tilde{B}$ works in
 both quantization schemes.


 \subsection  {The conserved quantum current}
 Just as in the classical case,  the current density can be
 constructed from the particle density by requiering the current to be
 conserved,
 \be{qcons}
 -\frac{i\bar k}{2} j_{ k} - \frac{ik}{2}\bar j_{ k}=
 \dot\rho_{ k} = \frac{1}{i\hbar}[\rho_k, H]\, ,
 \ee
 where $H$ is the full quantum Hamiltonian. From this we can derive
 two expressions for the current. The first is obtained  simply by
  evaluating the time derivative  of \pref{expanders}
 and noting that just
 as in the derivation of a classical current like \pref{infw}, a $\bar
 k$ or a $k$ can always be factored from the corresponding expressions,
 thus defining the complex components of the
 current. The explicit expressions, containing $\dot Z$ and $\dot
 Z^{\dagger}$,  are not very illuminating.
 Alternatively, we can assume some particular form for the Hamiltonian,
 evaluate the commutator,  and
 derive an explicit expression for the current in terms of the basic
 matrix operators. We examplify with a Hamiltonian consisting only of
 an external scalar potental, \ie
 \be{pot}
 H = V(\rho) = \int \frac {d^{2}k} {(2\pi)^{2}} V(k,\bar k )
 \rho_{ k} \, .
 \ee
 With this choice, \pref{qcons} takes the
 form
 \be{potcons}
 \dot\rho_{ k} = \frac{1}{i\hbar}  \int \frac {d^{2}p} {(2\pi)^{2}}\,
 V(p,{\bar p})[\rho_{ k},
 \rho_{ p} ]
 = \frac{i}{\hbar}
 \int \frac {d^{2}p} {(2\pi)^{2}}\, \sum_{n=1}^{\infty}
 \frac {(\ell^2/2)^{n} } {n{!}} V(p,{\bar p})
 [(\bar k p)^{n} - (k\bar p)^{n} ] \rho_{ k +  p} \, ,
 \ee
 where we used \pref{ccom} for the commutator and expanded the
 exponential and sine functions
 in a power series. Comparing with \pref{qcons} we can now directly
 read off the current,
 \be{qqur}
 j_{ k} = - \frac{2}{\hbar}\sum_{n=1}^{\infty} \frac {(\ell^2/2)^{n} }
 {n!}\, \bar k^{n-1}
  \int \frac {d^{2}p} {(2\pi)^{2}}\, p^{n} V( p,\bar p)
  \rho_{\vec k + \vec p} \, ,
 \ee
 which can be written more compactly in real space. A Fourier
 transformation gives,
 \be{qqur2}
 j(z,\bar z) = \frac{i}{\hbar} \sum_{n=1}^{\infty} \frac {(2\ell^2)^{n} }
 {n!} \,
 \partial_{z}^{n-1} [\rho(z,\bar z) \partial_{\bar z} V(z,\bar z) ] \, .
 \ee
 This expression was derived earlier by Martinez and
 Stone\cite{mart93} using a second quantized formalism. The above
 derivation  is more general in that it
 applies to any density operator satisfying the algebra \pref{ccom},
 and in particular to the matrix model density operator \pref{expanders}.


 \section{Summary and open problems}
 We have extended the  classical
 quantum Hall matrix models of Susskind and Polychronakos to include
 couplings to an external field.
 We gave a general prescription for the construction of a conserved current
 and noticed that it suffers from ordering ambiguities even at the
 classical
 level. Nevertheless, zero and low momentum observables can be reliably
 calculated using the infinite matrix model and in particular we showed
 that the ground state density, the quantum Hall response and
 the response to a weak and slowly varying magnetic field agree with
 what is  expected  for a fractional QH system.

 It is much harder to construct the correct charge and current operators
 in the quantized matrix model. 
 The difficulties are due to the complicated intermingling of matrix and quantum ordering. 
 In this connection, we
 also notice the difficulties encountered in a recent attempt by Karabali
 and Sakita to find good particle coordinates in the quantum matrix
 model\cite{kara01}. It turns out that the most natural guess, \ie defining
 the
 coordinates via the coherent states of the annihilation operator $Z_{ij}$,
 is not correct in that it does not reproduce the Laughlin wave functions.
 It is not unlikely that this is related to the failure of the anti-ordered
 matrix expression \pref{aoker} to satisfy the correct
 density operator algebra.

 Although not conclusive, our study supports the conjecture
 that the classical quantum Hall matrix model captures important
 aspects of the LLL physics not present in the usual effective
 Chern-Simons theory. At the practical level, the conclusion
 is, however, rather disappointing since local observables like
 charges and currents are rather complicated and thus cumbersome
 to work with. This might perhaps have been anticipated, since
 the gauge invariant objects in non-commutative theories are
 inherently nonlocal. On the other hand, it is the very fact that
 a non-commutative field theory  can be viewed as a
 commutative theory with an infinite number of higher
 derivative terms, that makes it at all possible that
  the Landau level structure could be incorporated at the
 level of an effective Lagrangian. 
 

 We end with a comment on the relation between the classical and
 the quantum models. It is striking that the {\em classical}
 matrix model incorporates  features
 of the FQHE which are related to the Landau level structure and
 not easily described in conventional mean field approaches.
  The $\theta$-dependent repulsion between the particles
 is  a phase space exclusion effect which can be thought of as the
 classical counterpart of fermion, or more generally anyon, statistics.
 In this way the classical matrix model can emulate a fermionic system
 with Haldane type pseudopotentals. In the case of a quadratic
 confining potential, this idea is made manifest via the mapping
 onto the Calogero model\cite{poly1}, and it would be interesting
 if a more general connection could be made between the QH matrix
 model and classical exclusion statistics\cite{hll2001}. The analogy
 between the classical matrix model and the QH system goes even
 further in that the classical counterpart to the density operator
 algebra \pref{ccom} can be interpreted as a quantum commutator algebra
 of fractionally charged quasi-holes\cite{hkp2002}. We hope to return
 to some of these question in the future.

 \vskip 3mm \noi {\bf Acknowledgement:} We thank Alexis Polychronakos
 for many useful comments and suggestions, and Dimitra Karabali for
 an illuminating discussion about reference \onlinecite{kara01}.


\begin{thebibliography}{99}

 \bibitem{zee} For a review, see \eg A. Zee, ``{\it Quantum Hall
 fluids}'' in Proc. of South African School of Physics, Tsitsikamma -
 94; Springer Verlag. [cond-mat/9501022].

 \bibitem{zhang92} See \eg the reviews by S. C. Zhang, Int. J. Mod. Phys.
 {\bf
 6}, 25 (1992)  and A. Lopez and E. Fradkin, in {\it "Composite Fermions"}
 ed. Heinonen, World Scientific (1998).

 \bibitem{jain1} J. K. Jain and R. K. Kamilla, in {\it "Composite
 Fermions"}
 ed. Heinonen, World Scientific (1998).

 \bibitem{recent} B I. Halperin, P. A. Lee and N. Read, Phys. Rev. {\bf
 B47}, 7312 (1993); G. Murthy and R. Shankar, in ``{\it "Composite
 Fermions"}''
 ed. Heinonen, World Scientific (1998)  and D-H. Lee,
 Phys. Rev. Lett. {\bf 80}, 4745 (1998).


 \bibitem{girvin} S. M. Girvin, A. H. MacDonald, and P. M. Platzman,
 Phys. Rev. {\bf B33}, 2481 (1986).

 \bibitem{suss1}
 L.~Susskind, hep-th/0101029.

 \bibitem{poly1} A.~P.~Polychronakos,
 JHEP {\bf 0104}, 011 (2001)[hep-th/0103013].

 \bibitem{hakarl1} T. H. Hansson and A. Karlhede, cond-mat/0109413.

 \bibitem{nair1}
 V.~P.~Nair and A.~P.~Polychronakos, Phys. Rev. Lett.
 {\bf 87}, 030403 (2001) [hep-th/0102181].


 \bibitem{suss2} S. Hellerman and L.~Susskind, hep-th/0107200.

 \bibitem{poly2} A.~P.~Polychronakos, JHEP {\bf 0106}, 070 (2001)
 [hep-th/0106011];


 \bibitem{mor} B. Morariu and A. P. Polychronakos, JHEP {\bf 0107}, 006
 (2001)
 [hep-th/0106072].

 \bibitem{heller} S. Hellerman and M. Van Raamsdonk, JHEP {\bf 0110}, 039
 (2001) [hep-th/0103179].

 \bibitem{halp1} S. H. Simon, A. Stern and B. I. Halperin,
 Phys.Rev. {\bf B54} : R11114 (1996) [cond-mat/9604103].

 \bibitem{harvey2001} J. A. Harvey, ``Koamba Lectures of Noncommutative
 Solitons and D/Branes'', hep-th/0102076; In {\em Proceedings of
 Strings 2001}, hep-th/0105242.

 \bibitem{mart93} J. Martinez and M. Stone, Int. J. Mod. Phys.
 {\bf B} Vol. 7, No. 26, 4389 (1993).



 \bibitem{kara01} D. Karabali and B. Sakita, Phys. Rev. {\bf B64}, 245316
 (2001)[hep-th/0106016].

 \bibitem{hll2001} T. H. Hansson, J.-M. Leinaas and U. Lindstr{\"o}m,
  Phys. Rev. {\bf E63}, 026102 (2001).

 \bibitem{hkp2002} T. H. Hansson, A. Karlhede and A. Polychronakos,
 unpublished.

 \end{thebibliography}
 \end{document}